# IMPROVING PSNR-BASED QUALITY METRICS PERFORMANCE FOR POINT CLOUD GEOMETRY

*Alireza Javaheri, Catarina Brites, Fernando Pereira, João Ascenso*

Instituto Superior Técnico – Universidade de Lisboa and Instituto de Telecomunicações
alireza.javaheri@lx.it.pt, catarina.brites@lx.it.pt, fp@lx.it.pt, joao.ascenso@lx.it.pt

**ABSTRACT**

An increased interest in immersive applications has drawn attention to emerging 3D imaging representation formats, notably light fields and point clouds (PCs). Nowadays, PCs are one of the most popular 3D media formats, due to recent developments in PC acquisition, namely with new depth sensors and signal processing algorithms. To obtain high fidelity 3D representations of visual scenes a huge amount of PC data is typically acquired, which demands efficient compression solutions. As in 2D media formats, the final perceived PC quality plays an importance role in the overall user experience and, thus, objective metrics capable to measure the PC quality in a reliable way are essential. In this context, this paper proposes and evaluates a set of objective quality metrics for the geometry component of PC data, which plays a very important role on the final perceived quality. Based on the popular PSNR PC geometry quality metric, novel improved PSNR-based metrics are proposed by exploiting the intrinsic PC characteristics and the rendering process that must occur before visualization. The experimental results show the superiority of the best proposed metrics over state-of-the-art, obtaining an improvement up to 32% in the Pearson correlation coefficient.

## 1. INTRODUCTION

Nowadays, point clouds are a relevant 3D representation format due to the advent of efficient point cloud (PC) acquisition, coding and rendering solutions. Cultural heritage, geographical information systems, real-time 3D telepresence and autonomous navigation are just a few examples of immersive multimedia applications that can benefit from the PC representation format. Since a massive amount of data has to be stored and transmitted when dealing with PCs, lossy compression methods have to be used, aiming to reduce the PC representation bitrate while keeping the data fidelity/quality as high as possible. In this context, in January 2017, MPEG issued a Call for Proposals on Point Cloud Compression (PCC) [1]. This call led to the development of two PC coding solutions, notably the so-called Geometry-based Point Cloud Compression (G-PCC) standard [2], for static and progressive acquired content, and the Video-based Point Cloud Compression (V-PCC) standard [3], for dynamic content. These two different landmark PC coding schemes produce decoded PCs with rather different characteristics and artifacts. Naturally, objective quality assessment metrics that can accurately evaluate the perceived quality, notably when PC data is compressed, are critical to improve the final Quality of Experience offered to the end-users.

Currently, there are several objective quality metrics available in the literature to automatically measure the PC geometry quality. One class of those objective metrics measure the degraded PC quality using point-based distances [4]. For every point in a degraded/original PC, the nearest neighbor is obtained in the original/degraded PC and the Haussdorff distance or the mean squared error (MSE) based distance is computed over all pairs of points; this type of metrics is usually referred to as point-to-point (Po2Po) metrics. The main disadvantage of this type of metrics is that it does not consider the fact that PC points represent a surface of an object(s) in the visual scene. To solve this issue, point-to-plane (Po2Pl) metrics [5] have been proposed, which model the underlying surface at each point as a plane; this plane is perpendicular to the normal vector at that point. Again, point-to-point distance of every point to its nearest neighbor is computed, which is then projected along the corresponding normal vector. This type of metrics results in smaller errors for the points that are closer to the surface. Currently, the MPEG adopted PSNR-based PC geometry quality metrics that use point-to-point MSE (D1) and point-to-plane MSE (D2) distances [6].

Another proposal to measure PC geometry quality is to estimate the similarity between the underlying surfaces associated to the points in the original and degraded PCs [7]. If planes are used, the cosine similarity between normal vectors at two corresponding points can be used to find the similarity. Another type of point-based distance considers the reconstructed surface, by proposing point-to-surface (also called point-to-mesh) distances [8]. In this case, a mesh is reconstructed on the reference PC and then, for each decoded point, the distance of the point to the reference mesh is computed.

Naturally, all objective metrics attempt to have PC quality scores as close as possible to the human opinion scores. In this context, subjective studies to collect the opinion scores for different scenarios and applications are needed [9][10][11], especially to perceptually model the artifacts caused by popular PC codecs, such as the MPEG G-PCC and V-PCC. The opinion scores are fundamental to evaluate the objective metrics performance, i.e. the subjective-objective correlation between these scores and predicted quality values. In [12], it is shown that available metrics fail to predict accurately the quality of octree compression-like distortions. In [13], a subjective-objective study also shows that state-of-the-art metrics do not perform well to assess compressed PCs quality in the presence of different types of data and decoding distortions. However, the MPEG PSNR D1 and D2 have shown the best overall performance in two separate studies [14][15], despite their lower overall subjective-objective correlation.

In this context, the main objective of this paper is to propose new quality metrics that measure the level of geometry degradation of PCs with different characteristics, e.g. content type, number of points and distribution of the points (sparse vs. dense). Due to the way that PCs are acquired and pre-processed (before coding), the intrinsic resolution, a measure of distance between points in the 3D space, plays an important role on the final perceived quality, not only to mitigate or highlight coding artifacts but also to measure the intrinsic PC quality (i.e. after acquisition). The main contributions of this paper are the following:

- Propose and evaluate several geometry quality PSNR-based metrics that exploit the *intrinsic characteristics* of a PC. In this case, intrinsic resolution and precision are considered the most important intrinsic characteristics that influence the final PC quality. The main novelty regards the new intrinsic resolution estimators (Section 3).
- Propose and evaluate a novel geometry quality PSNR-based metric that exploits the way that PCs are typically rendered. In this case, the intrinsic resolution is also considered but after rendering. This allows to significantly increase the quality metric performance, i.e.

to obtain higher correlation with human opinion scores.

The quality metrics that are proposed and evaluated keep the simplicity of the state-of-the-art Po2Po and Po2Pl metrics and are based on the popular PSNR metrics that are currently being used in the MPEG 3DG group to evaluate the performance of the MPEG G-PCC and V-PCC codecs and new techniques and are among the best.

The rest of this paper is organized as follows. Section 2 describes the state-of-the-art PSNR PC geometry quality metrics. Sections 3 and 4 present the proposed metrics. Experimental results are presented and analyzed in Section 4 and Section 5 concludes the paper.

## 2. PSNR-BASED GEOMETRY QUALITY METRICS

This section describes the state-of-the-art PSNR metrics as currently defined and used by the MPEG group [6] to assess the geometry quality for compressed PCs, independently of the codec used, target quality or rendering solution. As it will be seen in sections 3 and 4, the MPEG PSNR metrics are the starting point for the proposed PC geometry quality metrics.

In the MPEG PSNR-based geometry quality metrics, the PSNR is obtained from a normalization factor and the mean squared error (MSE), as defined in (1), which is computed from decoded to original PC direction as well as in the opposite direction. The PSNRs of the two directions are then combined to obtain a single symmetric PSNR value with the maximum pooling function, as defined in (2).

$$\text{PSNR}_{A,B} = 10 \log_{10}\left(\frac{p_s^2}{d_{A,B}^{MSE}}\right) \quad (1)$$

$$\text{PSNR} = \max(\text{PSNR}_{A,B}, \text{PSNR}_{B,A}) \quad (2)$$

In (1), $p_s$ is signal peak and $d_{A,B}^{MSE}$ is the average squared error (i.e. MSE) between all points in PC A and their corresponding nearest neighbor point in PC B. PCs A and B can be associated to original and decoded point clouds, respectively.

MPEG defines two PSNR metrics, the so-called PSNR D1 and PSNR D2, that only differ in way the MSE is computed. In the PSNR D1, MSE is computed from the point-to-point distance or error $\vec{e}(i,j)$ between each point in PC A and its nearest neighbor in PC B, $d_{A,B}^{Po2Po}$, according to (3) and (5); in (3), $j$ is the nearest neighbor point in PC B to point $i$ in PC A and $\vec{e}(i,j)$ is error vector between two corresponding points and $N_A$ in (5) stands for the number of points in PC A. In the PSNR D2, MSE is still based on the distance between each point to its nearest neighbor but this distance is now computed from the projection of the point-to-point error vector $\vec{e}(i,j)$ along the normal vector of the underlying surface at point $j$ in PC B, $\vec{n}_j$, as formalized in (4) and (5); the obtained distance, $d_{A,B}^{Po2Pl}$, is known as point-to-plane distance.

$$d_{A,B}^{Po2Po} = \|\vec{e}(i,j)\|_2^2 \quad (3)$$

$$d_{A,B}^{Po2Pl} = \|\hat{e}(i,j)\|_2^2 = (\vec{e}(i,j) \cdot \vec{n}_j)^2 \quad (4)$$

$$d_{A,B}^{MSE} = \frac{1}{N_A} \sum_{\forall a_i \in A} d_{A,B}^{Po2Po/Po2Pl}(i) \quad (5)$$

Thus, PSNR D1 can be obtained using (3) in (5) and (5) in (1) while PSNR D2 can be obtained using (4) in (5) and (5) in (1). As far as the signal peak $p_s$ in (1) is concerned, the largest diagonal (LD) distance of the PC bounding box is typically used for non-voxelized data [16]:

$$LD = \|(x_{\max}, y_{\max}, z_{\max}) - (x_{\min}, y_{\min}, z_{\min})\|_2 \quad (6)$$

where $x$, $y$ and $z$ are the three geometry coordinates. When the LD of the PC bounding box is used as peak in the PSNR computation, the point coordinates are normalized to the range [0,1], which is equivalent of having both original and decoded PCs fitted to a unit size bounding box. However, note that if the two PCs have different sizes they can be scaled differently; this scenario may occur when the decoded PC has errors with a large magnitude.

If PCs have been voxelized, the (point) coordinates lie on a regular 3D (integer) grid with some fixed predefined precision (coordinates bit-depth). Thus, point coordinates are bounded between zero and a constant integer related to the PC precision. For voxelized PCs, the peak for each coordinate, represented with $b$ bit-depth precision, is:

$$p_c = 2^b - 1. \quad (7)$$

By applying the peak of each coordinate in (7), the signal peak value corresponds to:

$$p_s = \sqrt{3} p_c. \quad (8)$$

By using (8) in (1), PCs are scaled to the same precision and the PSNR is computed as follows:

$$PSNR_{A,B} = 10 \log_{10}\left(\frac{3p_c^2}{d_{A,B}^{MSE}}\right) \quad (9)$$

Note that increasing the PC precision makes distances between points larger; for example, a PC with 11-bit precision has two times larger distances compared to the same PC with 10-bit precision. The normalization of the errors proposed in (9) makes the comparison of data in different precisions possible by compensating for this difference.

However, none of these metrics account for differences in: 1) intrinsic PC characteristics, namely the intrinsic resolution used to represent the surface of the objects of the PC, and 2) viewing conditions, namely the important rendering process that influences the final PC quality and can mitigate or highlight some coding distortions [14]. The PC spatial resolution is a key factor, since the perceivability of details depends on the sampling frequency, the distance of the PC to the image plane and from the image plane to the observer; these last two are assumed fixed in this work. In practice, independently of the coding artifacts, the subjective evaluation of a given PC varies when these factors vary.

## 3. INTRINSIC RESOLUTION PSNR-BASED QUALITY METRICS

In this section, besides the usual error to account for coding distortions, the PC intrinsic characteristics are also exploited to design improved PC geometry quality metrics.

In [5], the intrinsic resolution is used to normalize the geometric errors in the PSNR calculation (1) described in the previous section. The intrinsic resolution is difficult to measure since a point cloud is unstructured and the neighborhood of a point is more complex to define than that of a pixel in a 2D image. However, the PC intrinsic resolution concept is analogous to the spatial resolution of a 2D image and can be estimated from the distance of points to their neighbors [17]. This distance may not be uniform throughout the entire PC and requires the design of suitable estimators. Fig. 1 shows two PCs (scaled to the same height) with high (left) and low (right) intrinsic resolutions (i.e. dense and sparse), after point-based rendering with the same point size.

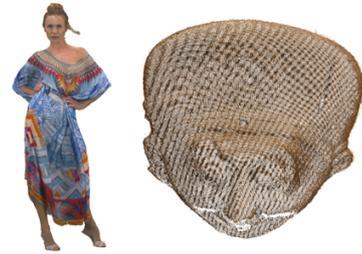

**Fig. 1**. Two PCs with different intrinsic resolutions: dense (left) and sparse (right).

Simple methods to estimate the PC intrinsic resolution were already proposed in the past, e.g. [5] suggests the maximum nearest neighbor ($MNN$) distance over all PC points. If $d_i$ is the distance of the point $i$ to its nearest neighbor in PC $O$, then the intrinsic resolution can be estimated according to:

$$MNN = \max_{\forall i \in O} d_i \quad (10)$$

The $MNN$ is very sensitive to holes and local sparse areas, even if they are very small comparing to the PC size. In this context, the following two estimators are proposed in this paper to overcome this problem. First, the average nearest neighbor ($ANN$) that attempts to solve the problem of localized sparse areas:

$$ANN = \sqrt{\frac{1}{N_O}\sum_{i \in O} d_i^2} \qquad (11)$$

where, $N_O$ is the number of points in the original PC $O$. Second, this metric can be generalized if the average of $k > 1$ nearest neighbors ($ANN_k$) is used instead of the closest nearest neighbor used in (11), which may not be very reliable due to acquisition noise. The $ANN_k$ can be defined as:

$$ANN_k = \sqrt{\frac{1}{N_O}\sum_{i \in O}\left(\frac{1}{k}\sum_{j=1}^{k} d_{i,j}^2\right)} \qquad (12)$$

Quality metrics can thus be defined by using the intrinsic resolution estimators proposed above (11)-(12) as a normalizer to convert the mean square errors to PSNR values. This means that the peak $p_s$ parameter in (1) will correspond to one of the intrinsic resolution estimators computed over the original PC (10)-(12), instead of the largest diagonal distance of the bounding box (6) or the precision established in the D1 and D2 metrics (8). In that way, PCs with different distances between points (i.e. more sparse or dense) would be normalized accordingly, i.e. all MSE errors would be scaled according to the estimated intrinsic resolution. However, all these metrics still do not consider the rendering process that is typically applied after decoding for visualization purposes.

## 4. RESOLUTION ADAPTIVE PSNR-BASED QUALITY METRICS

Point clouds are rendered as images or videos from one or more viewpoints before being shown on a 2D (or 3D) display, this means considering that some virtual observer is at some location in the 3D world space with some virtual camera setup that determines which parts of the point cloud are shown to the user.

In this process, the distances between all points in a PC are scaled with respect to this viewing box and display resolution. The viewing box is defined by the virtual camera position, orientation and characteristics (e.g. field of view) and the 3D to 2D projection (often a perspective projection is used). Since PCs are always evaluated by the users after rendering, the final perceived quality does not only depend on the PC errors introduced by some processing step (in this case, coding) but also on the rendering process. In this section, a novel PC geometry quality metric is proposed based on the idea of estimating the intrinsic resolution but after rendering, i.e. considering the distance of each point to its nearest neighbors on the image plane observable by the user. This is hereafter referred to as *rendering resolution* and a procedure to estimate it is described next.

### 4.1. Rendering Resolution Estimation

The rendering resolution may vary differently in different parts of the PC due to several factors, e.g. orientation of the PC surfaces in the 3D world relative to the observer viewing location. Since the viewing location and other rendering parameters, e.g. distance of the PC to the image plane, are not known when some geometric quality metric is computed, it is assumed that the PC is viewed from all possible directions and from a fixed distance. This is a usual scenario in many applications (e.g. cultural heritage), however further optimizations are possible such as using multiple distances (or scales) as other 2D quality metrics such as MS-SSIM [18].

Consider that a unit normal vector $\vec{n}_i$ is available for every $a_i$ PC point. Normal vectors, which are already used in point-to-plane PC quality metrics, can be quickly estimated using some state-of-the-art method. The rendering resolution estimation process is illustrated in Fig. 2 and proceeds as follows:

1. Define a local neighborhood $\gamma$ around point $a_i$, which includes the closest (nearest neighbors) $k$ points (represented by the six red dots in Fig. 2).
2. Define a local plane tangent at point $a_i$. This plane is perpendicular to the normal vector $\vec{n}_i$ and represents the PC surface at this point. This plane can also be seen as the image plane of the rendering process if the viewing direction (usually defined as a vector) is symmetric to $\vec{n}_i$ and the observer location lies along the viewing direction.
3. Project all points of the $\gamma$ neighborhood, represented by the $\vec{d}_{i,j}$ vectors in Fig. 2, to the local plane tangent at point $a_i$ according to (13); in (13), $\vec{d}_{i,j}$ stands for the distance vector between point $a_i$ and its $j$th nearest neighbor. This step results in a vector for each projected point (represented by the red dashed arrows in Fig. 2), called *planar distance vector* $\vec{PD}_{i,j}$, whose origin is $a_i$.

$$\vec{PD}_i = Proj_{plane}(\vec{d}_{i,j}) = \vec{d}_{i,j} - Proj_{\vec{n}_i}(\vec{d}_{i,j}) \qquad (13)$$

4. Estimate the rendering resolution as the average (planar) distance between point $a_i$ and their $k$ local neighbors on the tangent plane. This is performed for all points $a_i$ to obtain a global estimation for the entire point cloud. In this case, averaging over all points resulted in a better estimation. Thus, the rendering resolution corresponds to the average over the entire original PC of the average (planar) distance between point $a_i$ and the $k$ nearest neighbors in $\gamma$, $APD_k$, as formalized in (14).

$$APD_k = \frac{1}{N_O}\sum_{i \in O}\left(\frac{1}{k}\sum_{j=1}^{k}\|\vec{PD}_{i,j}\|_2^2\right) \qquad (14)$$

Since $APD_k$ includes the distance between points considering that some projection was performed in the rendering process, it is expected that it better reflects the intrinsic perceived PC quality when used in a geometry quality metric. The proposed resolution adaptive (RA) PSNR metric that exploits this factor is described next.

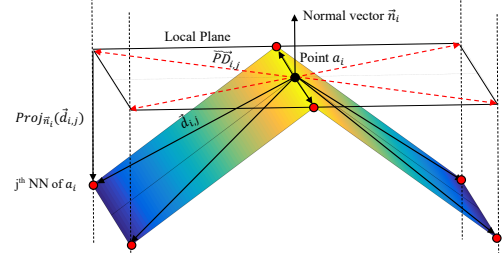

**Fig. 2.** Illustration of the proposed rendering resolution estimation process.

### 4.2. RA-PSNR PC Quality Metric Design

The proposed RA-PSNR objective metric aims to account for the PC intrinsic quality, by exploiting the intrinsic or rendering resolution as well as the precision used for the PC geometry coordinates. Ideally, the objective metric should compensate for any difference between PCs if one of these factors changes (e.g. precision) without having influence on the final perceived quality. For example, if the PC precision increases one bit without adding any new points, all the distances between points, used to calculate the intrinsic resolution, are now two times larger but the PC typically has a very similar perceived quality. In order to compensate for those factors, a density coefficient $\mu$ is defined as:

$$\mu = \frac{p_c}{r} \qquad (15)$$

where $p_c$ is the precision (coordinate peak) calculated with (7), i.e. the maximum possible distance, and $r$ is the PC intrinsic or rendering resolution. The density coefficient corresponds to normalization of the intrinsic or rendering resolution (which is typically defined as the inverse of density) according to the PC precision. Using the state-of-the-art PSNR D1 and D2 metrics [6] defined by (9), all the errors (which represent a distance between points in the original and degraded PCs) are normalized according to precision. Therefore, the intrinsic or rendering resolution of the PCs proposed in Section 3 and 4.1, which are also distances between points should also be normalized accordingly. The proposed resolution adaptive RA-PSNR is defined in (16) and corresponds to using the density coefficient $\mu$ to further scale the MSE, taking into account the intrinsic PC characteristics and the rendering process in the quality metric.

$$\text{RA-PSNR}_{A,B} = 10 \log_{10} \frac{3 p_c^2}{\mu d_{A,B}^{MSE}} \quad (16)$$

By applying (15) in (16), the RA-PSNR can be written as:

$$\text{RA-PSNR}_{A,B} = 10 \log_{10} \frac{3 r p_c}{d_{A,B}^{MSE}} \quad (17)$$

The parameter $r$ can be the intrinsic resolution from (10) to (12) or the rendering resolution from (14), computed from the original PC.

## 5. PERFORMANCE EVALUATION

This section evaluates the performance of the proposed PC quality metrics using point clouds coded with different codecs, notably the MPEG PC standard codecs.

### 5.1. Subjective Test Conditions

In these experiments, the MOS scores obtained in a previous subjective DSIS test described in [14] are used. The IST Rendering Point Cloud Dataset is publicly available in [19] and includes MOS scores and the original and decoded PCs. This dataset includes six PCs from the MPEG repository [6], which have been coded with three rates/qualities using three rather different PC codecs: i) octree-based codec from PCL [20]; ii) MPEG G-PCC standard [2]; and iii) MPEG V-PCC standard [3]. The quality scores were obtained for a session where the decoded PCs were rendered with a popular point-based rendering approach with uniform color and shading. Because there are no coloring and interpolation processes involved in this rendering, geometry coding artifacts are less masked. To evaluate the objective-subjective correlation performance of the proposed PC geometry quality metrics, non-linear regression has been used to map the computed objective quality scores (PSNR) into the MOS scale. Thus, based on the recommendation ITU-T P.1401[21], a monotonic cubic function was used to fit objective values to MOS scores and obtain predicted MOS:

$$\text{MOS}_p = \beta_1 + \beta_2 x + \beta_3 x^2 + \beta_4 x^4 \quad (18)$$

where $x$ is objective metric values and $\beta_1, \ldots, \beta_4$ are the regression model parameters. Then, the Pearson Linear Correlation Coefficient (PLCC) and Spearman Ranked Order Correlation Coefficient (SROCC) are used to assess the objective-subjective correlation, i.e. as measures of linearity and monotonicity of the prediction of perceptual quality made by the objective metrics. The following PC quality metrics were evaluated:

1. **MPEG PSNR D1 and D2 metrics:** These benchmark metrics represent the state-of-the-art and were described in Section 2. Two variants are defined, using precision ($P$) and the largest diagonal ($LD$) distance of the bounding box as the signal peak.
2. **Proposed I-PSNR metrics:** PSNR-based metrics where the $MNN$, $ANN$, $ANN_k$ intrinsic resolutions estimators (variants) described in Section 3 are used. The intrinsic resolution is used as the signal peak of the PSNR metric described in Section 2.
3. **Proposed RA-PSNR metrics:** PSNR-based metrics where the proposed intrinsic and rendering resolution estimators $ANN$, $ANN_k$ and $APD_k$ (variants) described in Section 3 and 4.1 are used. The RA-PSNR metric corresponds to (17) as presented in Section 4.2.

For all the experimental results shown, the $k$ parameter of $ANN_k$ and $APD_k$ was set to 10, which was experimentally found as the value that provides the highest Pearson correlation with MOS scores. For all metrics, the performance of the normalization factors $P, LD, MNN, ANN, ANN_k$ and $APD_k$ are shown for Po2Po and Po2Pl errors.

### 5.2. Results and Analysis

Table 1 shows the objective metrics performance for each PC codec and considering all data. The following conclusions can be made:

- *Proposed PC quality metrics vs benchmarks*: The reference benchmark PSNR metrics are outperformed with the proposed RA-PSNR or I-PSNR for all codecs individually and for all data. I-PSNR metrics can achieve higher performance than PSNR metrics for PCL and V-PCC while RA-PSNR is consistently better for all cases. The best overall performance is achieved for RA-PSNR with the rendering resolution $APD_k$ variant, which was expected since it exploits the rendering process required for PC visualization. Moreover, for the all case, where different types of artifacts and PC characteristics must be accounted, 5.3% gains were achieved.
- *Quality metrics performance for PCL and G-PCC data*: For PCL and G-PCC decoded data, the RA-PSNR proposed metric $APD_k$ variant can reach a very high performance (95.2% and 94.0%, respectively), since it considers both precision and rendering resolution, i.e. the point cloud density observed by the user.
- *Quality metrics performance for V-PCC data*: For V-PCC decoded data, I-PSNR $ANN_k$ variant performs slightly better than RA-PSNR $APD_k$, because all PCs are coded in 10-bit with V-PCC and the intrinsic resolution is already a good estimator of the quality after rendering. In a general way, the benchmark PSNR metrics are very poor for V-PCC data but both the proposed I-PSNR and RA-PSNR were able to increase performance significantly, with 28% and 26% gains respectively, compared to PSNR, for point-to-plane metrics.

TABLE I. PLCC PERFORMANCE OF THE PROPOSED PC OBJECTIVE METRICS

| Metric | Variant | Type | PCL | | G-PCC | | V-PCC | | All | |
|---|---|---|---|---|---|---|---|---|---|---|
| | | | PLCC | SROCC | PLCC | SROCC | PLCC | SROCC | PLC | SROCC |
| PSNR | P | Po2Po | 87.0 | 73.9 | 86.9 | 87.4 | 53.1 | 62.0 | 67.3 | 64.7 |
| | | Po2Pl | 89.6 | 80.9 | 83.4 | 85.6 | 51.4 | 49.6 | 70.3 | 65.6 |
| | LD | Po2Po | 83.3 | 82.3 | 86.0 | 89.3 | 48.9 | 54.1 | 70.4 | 68.6 |
| | | Po2Pl | 86.7 | 85.9 | 75.6 | 71.9 | 59.9 | 58.9 | 71.4 | 67.2 |
| I-PSNR | MNN | Po2Po | 68.7 | 65.0 | 40.8 | 33.3 | 45.6 | 14.3 | 49.7 | 42.2 |
| | | Po2Pl | 69.6 | 66.7 | 44.1 | 39.7 | 49.8 | 25.4 | 52.1 | 43.4 |
| | ANN | Po2Po | 92.2 | 89.0 | 79.3 | 76.3 | 66.6 | 61.0 | 64.7 | 52.5 |
| | | Po2Pl | 92.3 | 84.5 | 86.6 | 79.0 | 76.2 | 69.1 | 66.4 | 55.6 |
| | ANNk | Po2Po | 88.8 | 87.0 | 78.1 | 68.4 | 70.8 | 62.9 | 66.5 | 59.2 |
| | | Po2Pl | 90.8 | 87.5 | 75.3 | 68.2 | **79.6** | **74.0** | 67.4 | 62.6 |
| RA-PSNR | ANN | Po2Po | 92.8 | 86.8 | 84.9 | 82.5 | 49.2 | 45.6 | 68.9 | 64.0 |
| | | Po2Pl | 93.4 | 86.7 | 88.5 | 85.8 | 67.9 | 63.0 | 71.0 | 67.1 |
| | ANNk | Po2Po | 94.1 | 86.2 | 93.6 | 94.6 | 68.5 | 62.1 | 74.1 | 71.1 |
| | | Po2Pl | 95.1 | 91.3 | 94.0 | 94.0 | 77.1 | 73.7 | 74.9 | 72.5 |
| | APDk | Po2Po | 94.1 | 85.2 | 93.7 | **94.8** | 59.9 | 59.7 | 74.0 | 71.4 |
| | | Po2Pl | **95.2** | **91.7** | **94.0** | 94.0 | 77.9 | 72.3 | **75.6** | **73.8** |

## 6. CONCLUSIONS

State-of-the-art PC geometry quality assessment metrics perform poorly since the impact of the intrinsic PC characteristics and the rendering process on the final perceived PC quality are not considered. In this paper, the popular PSNR based metrics are improved by proposing to include a normalization factor that accounts for changes in the intrinsic PC resolution after rendering, as well as PC precision. Experimental results show that the proposed metrics outperform state-of-the-art MPEG quality metrics by a significant margin. As future work, more characteristics of the rendering process may be included, such as viewing location and PC distance to the viewing plane.